# Enhanced Optoelectronic Response in Bilayer Lateral Heterostructures of Transition Metal Dichalcogenides

Prasana K. Sahoo[1]†*, Shahriar Memaran[2,3]†, Yan Xin[2], Tania Díaz Márquez[1], Florence Ann Nugera[1], Zhengguang Lu[2,3], Wenkai Zheng[2,3], Nikolai D. Zhigadlo[4], Dmitry Smirnov[2,3], Luis Balicas[2,3],* and Humberto R. Gutiérrez[1],*

[1] Department of Physics, University of South Florida, Tampa, Florida 33620, USA

[2] National High Magnetic Field Laboratory, Florida State University, Tallahassee, FL 32310, USA

[3] Department of Physics, Florida State University, Tallahassee, Florida 32306, USA.

[4] Department of Chemistry and Biochemistry, University of Bern, 3012 Bern, Switzerland.

† These authors contributed equally to this work.

* Corresponding authors: E-mail: humberto3@usf.edu; balicas@magnet.fsu.edu; prasanasahoo@gmail.com

Abstract

**Two-dimensional lateral heterojunctions are basic components for low-power and flexible optoelectronics. In contrast to monolayers, devices based on few-layer lateral heterostructures could offer superior performance due to their lower susceptibility to environmental conditions. Here, we report the controlled synthesis of multi-junction bilayer lateral heterostructures based on $MoS_2$-$WS_2$ and $MoSe_2$-$WSe_2$, where the hetero-junctions are created via sequential lateral edge-epitaxy that happens simultaneously in both the first and the second layer. With respect to their monolayer counterparts, bilayer lateral heterostructures yield nearly one order of magnitude higher rectification currents. They also display a clear photovoltaic response, with short circuit currents ~$10^3$ times larger than those extracted from the monolayers, in addition to room-temperature electroluminescence. The superior performance of bilayer heterostructures significantly expands the functionalities of 2D crystals.**

The integration of distinct transition metal dichalcogenide (TMD) monolayers on the same substrate has seen tremendous progress in recent years *via* the use of chemical vapor deposition techniques [1-10]. It is possible to grow wafer size high quality films [1,2] of these compounds, or even vertical [4] and lateral heterostructures composed of monolayers with distinct band gaps [3-10]. To demonstrate the feasibility of various optoelectronic applications, ranging from highly performing sub-thermionic tunnel field-effect transistors [11] to photodetectors [12], to possible memory elements [13,14] was the underlying motivation behind these studies. More recently, we reported a rather simple chemical vapor deposition (CVD) method to create an undefined number of lateral heterojunctions with epitaxially grown monolayer domains of distinct binary TMD compounds and their ternary alloys [9]. There is a large variety of reported device performances for a given TMD material, this variability among reports could have its origin at different material qualities as well as the high sensitivity of monolayers to the environmental conditions. Substrate roughness and impurities, as well as gas adsorbates can introduce unwanted scattering mechanisms for the charge carriers that are detrimental to device performance. Monolayers are also very sensitive to humidity and aging. Encapsulation could minimize this problem improving device performance [15,16]. Another alternative is to fabricate devices based on few-layers crystals that have shown superior performance when compared to their monolayer counterparts [17-20]. In this sense, lateral heterostructures based on bilayers of TMDs materials are expected to be a more robust system, both in terms of chemical stability and a lower susceptibility to substrate quality or environmental conditions, while maintaining the potential for low-power and flexible optoelectronics offered by monolayers. Thicker lateral heterostructures also providing a longer photon vertical path through the junction, which should increase photon absorption and hence the probability of electron-hole pairs generation. Bilayers of TMD materials also present attractive and unique physical properties associated to their distinctive crystal symmetry; for instance, robust valley polarization and valley coherence [21], as well as the possibility of electrically tuning the valley magnetic moment [22]. Despite the potential advantages mentioned above, there have been no reports on lateral heterostructures based on bilayer or few-layers TMDs, mostly because they represent a greater challenge from the synthesis point of view. In the present work we demonstrate the synthesis of multi-junction lateral heterostructures of bilayer TMD domains, which exhibit a significantly superior optoelectronic response when compared to monolayers.

The bilayer heterostructures were synthesized directly on a $SiO_2$/Si substrate following our previously reported one-pot water-assisted chemical vapor deposition (CVD) approach [9] (for details see the methods section). Figures 1a, 1b and 1c, show optical images of bilayer $MoS_2$-$WS_2$ lateral heterostructures with one-, three- and seven-junctions, respectively. The sulfides-based heterostructures predominantly grew in truncated triangular shapes (supplementary information Fig. S1a). Figure 1d shows the schematic representation of a typical $2H_c$ type bilayer lateral heterostructure. The representative Raman spectra from $WS_2$ and $MoS_2$ individual domains are shown in Fig. 1e. The Raman peaks at 354 cm$^{-1}$ ($E_{2g}$ mode), 420 cm$^{-1}$ ($A_{1g}$ mode) and 350 cm$^{-1}$ ($2LA(M)$) are characteristic of $WS_2$ [23,24]; while, the peaks at 383 cm$^{-1}$ ($E_{2g}$) and 407 cm$^{-1}$ ($A_{1g}$) correspond to $MoS_2$ (Fig. 1e) [25]. The composite Raman intensity map in Fig. 1f clearly shows the spatial distribution of laterally connected domains of $MoS_2$ (407 cm$^{-1}$) and $WS_2$ (350 cm$^{-1}$) bilayers (see additional maps in supplementary information Fig. S1).

The photoluminescence (PL) spectrum from 2L-$WS_2$ bilayer domains (Fig. 1g) consists of one major peak located at $\cong$ 645 nm accompanying a minor broad peak at $\cong$ 720 nm. The peak at 645 nm corresponds to direct excitonic emission at the K and K' points of the Brillouin zone. Whereas the peak at 720 nm arises from indirect electronic transition. Even though the electronic bands in $WS_2$ evolve from a direct to an indirect band gap semiconductor as the number of layers increases [26], bilayer $WS_2$ still displays a relatively strong direct band gap emission that dominates its PL spectrum. For the 2L-$MoS_2$ bilayer domains (Fig. 1g) the typical PL spectrum presents two major peaks around 680 nm and 606 nm, corresponding to the A and B excitonic transitions, respectively [27]. The PL spectrum at the WS$_2$→MoS$_2$ hetero-interface (Fig. 1g) consists of a superposition of PL peaks from both MoS$_2$ and WS$_2$ domains that are simultaneously excited by the laser probe. The absence of alloy-related peaks at intermediate energies suggests the presence of a chemically sharp WS$_2$→MoS$_2$ interface. The PL intensity map as well as the peak position map (in Figs. 1h and 1i, respectively) are in agreement with the Raman map in Fig. 1f, (additional maps in supplementary information Fig. S1d). The normalized PL contour plot of a line-scan across three-junctions (Fig. 1j) better visualizes the abrupt modulation of the band gap. Notice that the transition of the PL signal for WS$_2$→MoS$_2$ interfaces is sharper when compared to MoS$_2$→WS$_2$ interfaces, the latter might present a small degree of alloying at the junction. The existence of hetero-junctions with different composition profile (sharp or smooth) is intrinsically related to the distinct kinetics of the chemical reactions of Mo- and W-based compounds during the gas switching time. We previously observed a similar trend in monolayer lateral heterostructures, for a detailed discussion on the growth mechanisms we refer the reader to our previous report in ref. [9].

The atomic structure and the local chemical distribution near the bilayer lateral heterojunctions were studied by atomic resolution Z-contrast imaging and by Electron Energy Loss Spectroscopy (EELS) in an aberration-corrected Scanning Transmission Electron Microscope (STEM). Figure 1k shows a low magnification, high-angle annular dark field (HAADF) STEM image of a heterostructure section, the 2L-WS$_2$ domain appears in bright contrast while the 2L-MoS$_2$ domain is darker. An atomically-sharp bilayer lateral heterojunction is shown in Fig. 1l, where the atoms appear brighter in the WS$_2$ domain due to the higher Z-number. The crystal structure in both domains (2L-WS$_2$ and 2L-MoS$_2$) display a hexagonal arrangement (Fig. 1l) that is consistent with the *2H$_c$* polytype [28], space-group $P6_3/mmc$. In the *2H$_c$* polytype each transition metal atom, in one layer, is vertically aligned with two chalcogen atoms located in the neighboring layer, and vice versa (see schematic in Fig. 1m and its corresponding intensity profile). Lower magnification Z-contrast images from a bilayer lateral heterojunction (Fig. 1n) as well as elemental EELS maps of the Mo-L and W-L edges (Fig. 1o) also indicate a clear difference in the chemical composition of each domain, which is consistent with the PL and Raman maps discussed above. Figures 1p and 1q show a thickness transition at the outermost edge of the heterostructure, from bilayer *2H$_c$* WS$_2$ to monolayer *1H* WS$_2$, and its corresponding scattered electron intensity profile along the $[11\bar{2}0]$ direction. The distinct symmetry of the atomic contrast between the bilayers and the monolayers, as observed in the STEM images, facilitates their identification.

Similar to the sulfides, bilayer lateral heterostructures of selenium based TMDs (2L-MoSe$_2$ – 2L-WSe$_2$) were also synthesized using the water assisted CVD method [9]. Figure 2a displays a three-junction heterostructure with faceted hexagonal geometry; the 2L-MoSe$_2$ domains appear in a reddish brown color, while 2L-WSe$_2$ domains are pink (see additional optical images in supplementary information Figs. S2 and S3). Representative Raman spectra for bilayer (2L) and monolayer WSe$_2$ and MoSe$_2$ domains are shown in Figs. 2b and 2c, respectively ( and supporting information Fig. S4). For the bilayer WSe$_2$ the *A$_{1g}$* phonon mode (at 250.5 cm$^{-1}$) is blue shifted by 1 cm$^{-1}$ when compared to the monolayer, while the peak at 257.5

cm$^{-1}$ (*2LA(M)*) is red shifted by 2.5 cm$^{-1}$. The frequency difference *Δω* [*2LA(M)* - *A$_{1g}$*] is 7 cm$^{-1}$ for bilayers and 11 cm$^{-1}$ for monolayers, in agreement with previous reports [29]. In bilayer MoSe$_2$, the characteristic phonon modes (Fig. 2c) were observed at 240 cm$^{-1}$ (*A$_{1g}$* mode) and at 249.5 cm$^{-1}$ ($E_{2g}^2$(*M*) shear mode). In contrast to bilayer WSe$_2$, the difference in frequency Δω [$E_{2g}^2$(*M*) - *A$_{1g}$*] for bilayer MoSe$_2$ is negligible. However, the intensity ratio A$_{1g}$/$E_{2g}^2$(*M*) decreases by 76 % in bilayer MoSe$_2$ when compared to its monolayer. Figure 2d displays the composite Raman intensity map at frequencies of 250 cm$^{-1}$ (2L-WSe$_2$) and 240 cm$^{-1}$ (2L-MoSe$_2$), the color pattern agrees with the optical contrast in Fig. 2a.

We observed a clear dependence of the PL spectra with the number of layers in both WSe$_2$ and MoSe$_2$ domains as shown in Fig. 2e (see also supplementary information Fig. S2), in agreement with previous reports [29-31]. In bilayer WSe$_2$, the PL peak corresponding to the direct excitonic emission appears at 788 nm, which is shifted by *ΔE = E$_{1L}$ - E$_{2L}$* = 8.6 meV when compared to the monolayer. The shoulder at ~811 nm is associated with the indirect bandgap optical transition. Similarly, for bilayer MoSe$_2$, the direct and indirect band transitions are located at 827 nm and 893 nm, respectively [31], with a direct bandgap shift of *ΔE = E$_{1L}$ - E$_{2L}$* = 9 meV with respect to the monolayer. The PL signal from the bilayer domains (WSe$_2$ and MoSe$_2$) is rather broad and exhibits relatively strong emissions, typical of high quality materials. Composite PL intensity maps (Fig. 2f) at 788 nm (2L-WSe$_2$) and 827 nm (2L-MoSe$_2$) further confirms the homogeneity of the PL signal within each domain (see also supplementary information Fig. S4). A PL line-scan perpendicular to the three-junctions (Fig. 1g) also confirms the sequential modulation of the bandgap across the heterostructure. Note that, similar to our observations in Fig. 1j for the sulfides-based heterostructures, the change in the position of the PL peak at the WSe$_2$→MoSe$_2$ junction is sharp when compared to the MoSe$_2$→WSe$_2$ interfaces (Fig. 1g). For the latter, the smooth change in the PL peak position across the interface suggests the presence of a compositional gradient, which contrasts with the sharp chemical transition observed for the WSe$_2$→MoSe$_2$ junctions. This trend does not depend on the type of chalcogen atom, but, as mentioned before, on the different oxidation-reduction kinetics of the Mo and W precursors [9].

A low magnification HAADF-STEM image of a bilayer (2L) WSe$_2$-MoSe$_2$-WSe$_2$ lateral heterostructure is shown in Fig. 2h. The 2L-WSe$_2$ domain appears in bright electronic contrast while the 2L-MoSe$_2$ domain is darker. The spatial chemical distribution observed from Z-contrast images (Fig. 2i), as well as elemental EELS maps (Fig. 2j and 2k) also confirm the formation of laterally connected 2L-WSe$_2$ and 2L-MoSe$_2$ domains, in agreement with the PL and Raman data shown in Figs. 2b-2g. Similar to the case of sulfides, there is a 1L-WSe$_2$ fringe domain at the outermost border of the heterostructure, presumably formed due to a reduction in precursors supply at the end of the synthesis process during the sample cooling (Fig. 2h). Atomic resolution imaging of this 2L-1L transition region (Fig. 2l) shows the distinct crystal symmetries of the 2L-WSe$_2$ and the 1L-WSe$_2$ domains. In contrast with the *2H$_c$* stacking observed in the sulfide bilayers, the symmetry observed for the selenides bilayers is more consistent with a *2H$_b$* stacking [28], space-group $P\bar{6}m2$, as depicted with the ball model in Fig. 2m.

In order to evaluate the optoelectronic properties of the bilayer lateral heterostructures, a particular configuration of metallic contacts have been designed to probe the electrical transport of the individual domains, as well as the transport properties across their junctions (Fig. 3a). The bilayer heterostructure in Fig. 3a is composed of three concentric triangular domains ($MoSe_2$ - $WSe_2$ - $MoSe_2$), and their junctions are highlighted by the red dashed lines. *I-V* characteristics measured across the junctions (*i.e.* across contacts 1-2 or 4-5, black symbols in Figs. 3b and 3c, respectively), show a clear diode-like response. The dark currents $I_{ds} \approx 200$ nA measured in forward bias ($V_{ds} > 0$) and $I_{ds} \approx 3.5$ nA measured in reverse bias (-1.5 V < $V_{ds}$ < 0), result in a $10^2$ ration between forward and reverse biases. Under illumination conditions the $I_{ds}$ in direct bias increases by a factor of three. Similarly to the monolayers [9] this marked non-linearity cannot be attributed to Schottky barriers at the level of contacts given the nearly linear *I-V* characteristics of individual domains, see Fig. 3d. The doping type for each individual domain, p-type $WSe_2$ and n-type $MoSe_2$, was determined by measuring the drain-source current ($I_{ds}$) as a function of back-gate voltage $V_{bg}$ (Fig. S5a).

The *I-V* characteristics across the junctions, shown in magnified scales in the insets of Figs. 3b and 3c, reveal the existence of a finite current under zero bias which scales with the illumination power, this photovoltaic-effect was not previously observed by us in the corresponding monolayer heterostructures [9]. Conventional solar cells are typically vertical PN-junctions, and the photovoltaic power conversion efficiency ($\eta$) is calculated dividing the maximum photogenerated electrical power $P = I_{ds} \times V_{ds}$ by the illumination power, vertical PN-junctions cover the entire illuminated area since they are parallel, regardless the sharpness of the junction in the vertical direction. Notice that for lateral junctions, the area of the junction is considerably small compared to the total illuminated area. As shown in Fig. 1l, the lateral junctions can be atomically sharp extending by a distance of less than 1 nm at the interface between both domains, and are only two monolayers in thickness (1.2 nm). Thus, taking in to account the laser spot size of 10 $\mu$m (diameter), regardless whether we consider the cross-sectional area or the in-plane area of the junction, the calculated area will be $A_j \cong 10^{-2}$ $\mu m^2$. Therefore, a laser illumination power of 1 $\mu$W would provide only $1.27 \times 10^{-4}$ $\mu$W to the junction area and the calculated efficiencies will be extremely high. To address this point, we fabricated contacts that expose distinct areas of the material surrounding the junctions, observing that the short circuit current or $I_{sc} = I_{ds}$ (when $V_{ds} = 0$ V) seems to scale with the illuminated area. In effect, we estimate illuminated areas of ~ 21.6 $\mu m^2$ and 10.8 $\mu m^2$ between contacts 1-2 and 4-5, respectively; with corresponding $I_{sc}$ values of $\cong$ 3 nA and 9 nA under 9 $\mu$W of laser illumination power ($\lambda$ = 532 nm). Therefore, these areas receive effective illumination powers $P$ = 2.48 $\mu$W and 1.23 $\mu$W, respectively; which can be contrasted to the extracted maxima photovoltaic electrical power $P_{el}^{max}$ = 2 nW and 0.4 nW, as indicated by the red dots in Figs. 3e and 3f, respectively. These numbers would lead to modest power conversion efficiencies $\eta$ = 100 x $P_{el}^{max}/P \cong$ 0.08 % and 0.03 %, respectively. Notice that if we use the previously estimated value of $A_j$ to calculate $\eta$, one would obtain values that are ~$10^3$ higher. Therefore, future work should focus in defining, unambiguously, the effective width of the junctions in lateral heterostructures as the ones studied here. However, small $\eta$ can be attributed to a number of factors, like the detrimental role played by the Schottky barriers, or the relatively small probability of generating electron-hole pairs by photons traveling through quite thin layers (low optical absorption). Nevertheless, one can envision a number of strategies to improve the conversion efficiency, i.e. use of reflective back layers, metals with distinct work functions for the electrical contacts, or the addition of extra layers to increase the efficiency of these heterostructures [32]. Figures 3g and 3h display $I_{sc}$, the open-circuit voltage $V_{oc}$, $P_{el}^{max}$, and the photovoltaic fill factor $FF = P_{el}^{max}/(I_{sc} \times V_{oc})$ as functions of the illumination power $P$. $I_{sc}$ and $P_{el}^{max}$ are observed to display a power law dependence on $P$ as previously observed in multilayered junctions [33]. $V_{oc}$ displays the typical semi-logarithmic dependence on $P$ while $FF$ shows very modest values, i.e. between 0.2 and 0.3, which are considerably smaller than the values extracted from conventional solar cells [32] indicating that there is ample room for improvement in performance.

The contacts configuration employed to study the electrical properties of lateral heterojunctions composed of $MoS_2$ and $WS_2$ bilayer domains is shown in Fig. 4a. Similar to the selenides, the individual domains display a p-type behavior for the $WS_2$ and n-type for $MoS_2$ (Fig. S5b). A clear diode like response is observed for the *I-V* characteristic measured between contacts 1-4, *i.e.* from the central $MoS_2$ bilayer domain towards the $WS_2$ outer domain (Fig. 4b). There is a pronounced photo-induced increase in the forward bias current. The $MoS_2$-$WS_2$ bilayers lateral junctions also display a clear photovoltaic response which is comparable in magnitude to that observed for $MoSe_2$-$WSe_2$ heterojunctions, the corresponding short-circuit currents are shown in the inset of Fig. 4b. The *I-V* characteristic corresponding to the $WS_2$ single domain (Fig. 4c), measured through contacts 5-6, is non-linear, indicating a more prominent role for the Schottky barriers around the contacts when compared to the Se based heterostructures. However, it does not display a clear diode-like response, the currents are 3 orders of magnitude inferior to the forward biased currents observed across the junction, and the effect of illumination is barely observable. This clearly indicates that the Schottky barriers play little to no role in the diode like response observed in $MoS_2$-$WS_2$ bilayer heterojunctions. The photo-generated electrical power is plotted in Fig. 4d. The illuminated area between the electrical contacts is 16 $\mu m^2$, which represents ~1/5 of the total laser spot area, hence 1 $\mu W$ in laser power actually corresponds to an incident power $P = 0.2$ $\mu W$. According to Fig. 4d, this yields a photo-generated $P_{el}^{max} = 0.043$ nW and thus an efficiency $\eta = 100 \times P_{el}^{max}/P = 0.02$ % which is comparable to the values previously extracted for the Se based heterojunctions. In these calculations we have assumed conservatively that the entire area of the exposed channel contributes to the photovoltaic response, but this remains to be confirmed. Figures 4e and 4f display $I_{sc}$, $V_{oc}$, $P_{el}^{max}$ and $FF$ as functions of the incident power $P$ in logarithmic scale. As previously seen for the Se based bilayer heterojunctions, both $I_{sc}$ and $P_{el}^{max}$ follow a power law dependence on $P$, with $V_{oc}$ displaying the canonical semi-logarithmic dependence on $P$. $FF$ on the other hand displays lower values, i.e. below 0.21, indicating a slightly inferior performance, possibly resulting from larger Schottky barriers as implied by the non-linear *I-V* characteristics of the individual domains.

Finally, the devices based on bilayer MoS$_2$-WS$_2$ lateral heterostructures exhibit electroluminescence (EL) at room temperature as shown in Fig. 4g. The electroluminescence signal is observed for bias voltages between 1.5 V and 2.5 V. The position of the EL peak maximum is between 1.91-1.94 eV, which is similar to the 1.92 eV direct excitonic emission in the PL spectrum of 2L-WS$_2$ domains (Fig.1j). Interestingly, these EL energy values are also comparable to the average energy (1.94 eV) between the most prominent peaks, excitons A and B, in the PL spectrum of the 2L-MoS$_2$ domains (Fig.1j). We have observed that the PL signal is considerably stronger in WS$_2$ than MoS$_2$, this suggests that the radiative electron-hole recombination process is more efficient in WS$_2$ and hence that the EL signal could be generated mainly in the 2L-WS$_2$ side of the junction. However, the EL peak is asymmetric towards lower energies, therefore a small contribution to the EL signal from electron-hole recombination (exciton A) in the 2L-MoS2 domain cannot be ruled out. Finally, notice that 1.94 eV corresponds to photons in the visible spectra, of red color approaching orange, which makes these junctions potential candidates for developing light emitting devices.

The ability to produce bilayer lateral heterostructures of transition metal dichalcogenides, through a newly developed and relatively simple chemical vapor deposition technique, offers an additional degree of freedom to create complex 2D device architectures and geometries. We demonstrated that bilayer lateral heterostructures have a superior optoelectronic response when compared to their monolayer counterparts, for instance, displaying clear photovoltaic and room-temperature electroluminescent responses. Although, the competition between the direct and the indirect electronic transitions in bilayer TMDs is detrimental to their photoluminescence quantum efficiency, the addition of an extra layer in 2D lateral heterostructures seems to minimize the role of substrates and adsorbates that could act as additional scattering centers for the charge carriers. The latter effect appear to be dominant, leading to a remarkable enhancement in device performance. Our observations suggest a promising route to produce more robust and reliable optoelectronic components based on few-layers transition metal dichalcogenides that are less sensitive to environmental factors.

**Methods**

**Synthesis of bilayer lateral heterostructures.** All in-plane bilayer lateral heterostructures were synthesized using the one-pot CVD approach recently developed by our group that is described in detail in the reference (*9*) of the main text. In brief, this method uses water-assisted thermal evaporation of solid sources at atmospheric pressure. Bulk powders of MoSe$_2$ (99.9%, Sigma Aldrich) and WSe$_2$ (99.9%, Sigma Aldrich) were placed together within a high purity Alumina boat and used as solid source precursors to synthesize the bilayer MoSe$_2$–WSe$_2$ heterostructures; while MoS$_2$ (99.9%, Sigma Aldrich) and WS$_2$ (99.9%, Sigma Aldrich) were the solid sources for bilayer MoS$_2$-WS$_2$ heterostructures. In each case, the powder sources containing 120 mg of MoX$_2$ and 60 mg of WX$_2$ [where X=S, Se] in a ratio of 2:1, were placed side-by-side within an alumina boat (L x W x H: 70 x 14 x 10 mm) in the center of a 1" diameter horizontal quartz tube furnace. SiO$_2$/Si (300nm oxide thickness) substrates were pre-cleaned with acetone, isopropanol and deionized water. During the growth, the substrates were placed downstream at temperatures ranging between 810 °C and 780 °C and 6-7 cm away from the solid sources which were maintained at 1060 °C. Initially, the temperature of the furnace was slowly raised up to 1060 °C within 50 min with a constant flow of $N_2$ (200 sccm) and both, substrates and sources, were kept outside the furnace. When the temperature of the furnace reached above 1040 °C, the solid precursor as well as the substrates were placed at their respective positions, by sliding the quartz tube into the furnace. Simultaneously, water vapor was introduced in a controlled manner by diverting $N_2$ flow through a bubbler (Sigma Aldrich) containing 2 ml of DI water at room temperature. In order to switch the growth from Mo- to W-rich compounds, the $N_2+H_2O$ vapor flux was suddenly replaced by a mixture of Ar + 5% $H_2$ (200 sccm). Multiple TMD domains were grown by sequentially switching the carrier gases: $N_2+H_2O$ vapor favors the growth of $MoX_2$ domains whereas changing the carrier gas from $N_2+H_2O$ to $Ar+H_2$ (5%) favors the growth of $WX_2$ (where X = S or Se). The lateral dimensions of individual TMD domains were independently controlled by varying the carrier gas flow times. We observed that increasing the amount of solid precursors and hence the deposition rate, promotes the formation of "bilayer" domains (2L $MoX_2$ – 2L $WX_2$) laterally connected via edge epitaxy. Similar growth conditions were employed for the growth of the selenium and the sulfur based heterostructures. Once the desired heterostructure sequence was completed, the synthesis process was abruptly terminated by sliding the quartz tube containing both the precursors and the substrates to a cooler zone, while keeping a constant 200 sccm flow of Ar + $H_2$ (5%) until it cooled down to room temperature.

**Raman and Photoluminescence Characterization.** The Raman and photoluminescense experiments were performed in a confocal microscope-based Raman spectrometer (LabRAM HR Evolution, Horiba Scientific) using a backscattering geometry. An excitation wavelength of 532 nm (laser power at the sample 77 μW) was used, and focused with a 100x objective (NA = 0.9, WD = 0.21 mm). During the PL and the Raman mapping the optical path is stationary, while moving the sample on a computer controlled motorized XY stage.

**Transmission Electron Microscopy.** High angle annular scanning transmission electron microscopy (HAADF-STEM) imaging was performed with an aberration-corrected JEOL JEM-ARM200cF with a cold-field emission gun at 200kV. The STEM resolution of the microscope is 0.78 Å. The HAADF-STEM images were collected with the JEOL HAADF detector using the following experimental conditions: probe size 7c, condenser lens (CL) aperture 30 μm, scan speed 32 μs/pixel, and camera length 8 cm, which corresponds to a probe convergence angle of 21 mrad and inner collection angle of 46 mrad.

**Device fabrication.** The electrical contacts to individual $MoX_2$ and $WX_2$ domains were fabricated by depositing 80 nm of Au onto a 8 nm thick layer of Ti *via* e-beam evaporation. Contacts were patterned using standard e-beam lithography techniques. After gold deposition, and in order to extract adsorbates, the samples were annealed under high vacuum for 24 hours at 120 °C. In order to access the inner domains in the heterostructures without short circuiting with the external ones, as in the case of $WSe_2$- $MoSe_2$ heterojunctions, thin *h*-BN crystals were mechanically exfoliated from larger crystals and conveniently placed on the heterostructure before depositing the contacts. The h-BN crystals were grown following the technique described in ref. (35). The technique used to transfer the *h*-BN onto the heterostructure is similar to the one described by Lee *et al.* [34].

**Electrical characterization** was performed using a sourcemeter (Keithley 2612 A). For photo-current measurements a Coherent Sapphire 532-150 CW CDRH and Thorlabs DLS146-101S were used, with a continuous wavelength λ = 532 nm. Light was transmitted to the sample through a 10 μm single-mode optical-fiber with a mode field diameter of 10 μm. The size of the laser spot was also measured against a fine grid.

**Acknowledgements.** HRG acknowledges support by the National Science Foundation Grant DMR-1557434 (CAREER: Two-Dimensional Heterostructures Based on Transition Metal Dichalcogenides). LB acknowledges support from Army Research Office through MURI award W911NF-11-1-0362 (Synthesis of 2D materials), the Office of Naval Research through DURIP Grant# 11997003 (transfer and stacking under inert conditions) and the National Science Foundation through DMR- 1807969 (synthesis and graduate student support). TEM work was performed at the NHMFL which is supported by the NSF Cooperative Agreement No. DMR-1644779 and the State of Florida.

**Author Contributions.** P.K.S., L.B. and H.R.G. conceived the idea, designed the experiments and wrote the manuscript. P.K.S. performed the CVD synthesis, Raman and Photoluminescence characterizations and related analysis; as well as sample preparation for TEM. T. D.M. and F. A. N. assisted the CVD growth and films transfer. Y.X. and H.R.G. conducted TEM experiments and related data analysis. S.M. and L.B. performed the device design and fabrication, electrical measurements and related analysis. W. Z. assisted the device fabrication and testing. Z.L. and D.S. contributed to the electroluminescence measurements. N. D. Z. grew the h-BN material used in the devices**.** All authors discussed the results and commented on the manuscript.

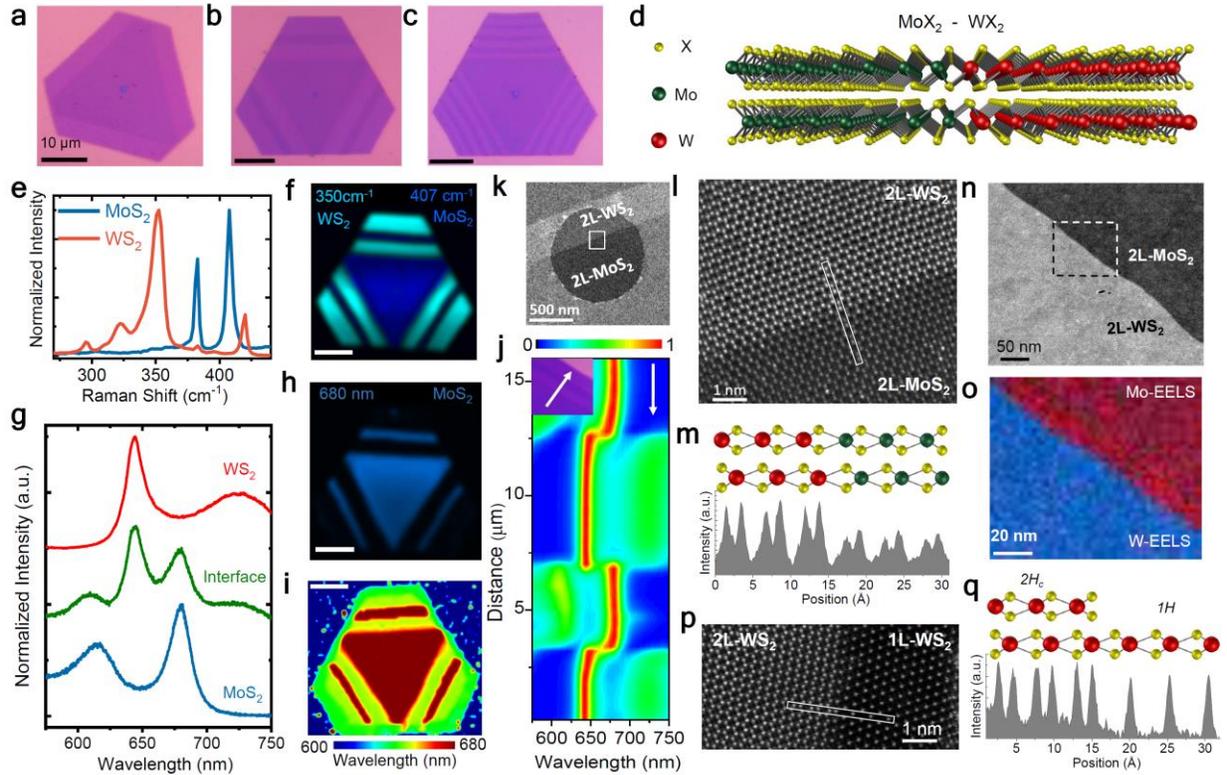

**Figure 1. Bilayer MoS$_2$-WS$_2$ lateral multi-junction heterostructure and their interfaces. a**, **b**, **c**, Optical images of bilayer MoS$_2$-WS$_2$ lateral heterostructures with single junction, three junctions and seven-junctions (periodic structure), respectively. **d,** Atomic structure of the bilayer heterostructure from a cross-section perspective, with 2*H* stacking. **e**, Typical Raman spectra of bilayer WS$_2$ and MoS$_2$ domains. **f,** Composite Raman intensity map including the modes at 405 cm$^{-1}$ (*A$_{1g}$* - MoS$_2$) and 350 cm$^{-1}$ (*2LA(M)* - WS$_2$) for the heterostructure in (**b**). **g,** Typical photoluminescence (PL) spectra of a 2L-WS$_2$ domain, a MoS$_2$ domain and the interface WS$_2$-MoS$_2$. **h**, PL intensity maps of MoS$_2$ domains (peak at 680 nm) for the structure in (**b**). **i**, PL position map for the peak maxima, corresponding to the structure in (**b**). **j**, Contour color plots of the normalized PL intensity as a function of the position across the three-junctions (for the structure in **b**) where the white arrows indicate the direction of the line map. Scale bars correspond to 10 μm. **k**, HAADF-STEM image of a lateral heterostructure section. **l**, Atomic-resolution STEM image of an atomically-sharp bilayer lateral heterojunction. **m**, Scattered electron intensity profile (along the line in (**l**)) and its corresponding ball model from a cross-sectional perspective. **n**, Lower magnification Z-contrast images from a bilayer lateral heterojunction and (**o**), its corresponding EELS maps of the Mo-L and W-L edges. **p**, atomic-resolution STEM image of a thickness transition from bilayer *2H$_c$* WS$_2$ to monolayer *1H* WS$_2$. **q**, Scattered electron intensity profile (along the line in (**p**)) and its corresponding ball model of the cross-section perspective.

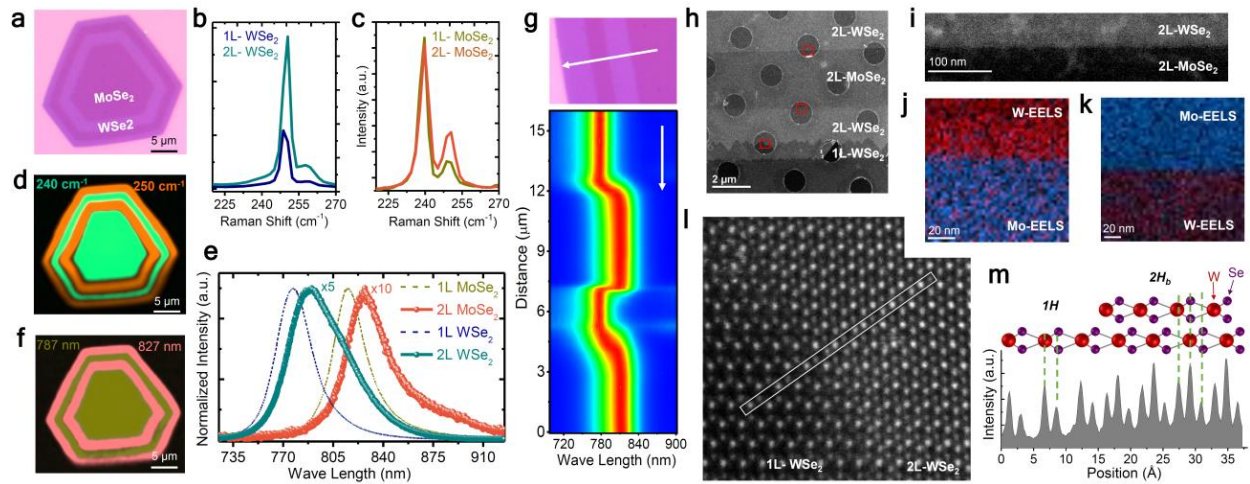

**Figure 2. Bilayer MoSe$_2$-WSe$_2$ lateral multijunction heterostructures and interfaces. a**, Optical image of a three-junction bilayer MoSe$_2$ - WSe$_2$ lateral heterostructure. **b**, Raman spectra of 1L- and 2L-WSe$_2$ domains. **c**, Raman spectra of 1L- and 2L-MoSe$_2$ domains. **d**, Composite Raman intensity map at the frequency 250 cm$^{-1}$ (A$_{1g}$ mode of 2L-WSe$_2$) and at 240 cm$^{-1}$ (A$_{1g}$ mode of 2L-MoSe$_2$). **e**, Normalized PL spectra of 1L- and 2L-WSe$_2$ as well as 1L- and 2L-MoSe$_2$ domains. **f**, Composite PL intensity map using the 788 nm (2L-WSe$_2$) and 827 nm (2L-MoSe$_2$) peaks. **g**, Top panel: optical image (section) of a bilayer MoSe$_2$-WSe$_2$ lateral heterostructure. Lower panel: normalized PL color contour plot across the white line in the top panel. **h**, Low magnification HAADF-STEM image of a bilayer (2L) WSe$_2$-MoSe$_2$-WSe$_2$ lateral heterostructure. **i**, Higher magnification Z-contrast image of the junction at the top red square in (**h**). **j, k,** Elemental EELS maps for the top and middle bilayer lateral junctions (red squares in (**h**)), respectively. **l**, Atomic resolution STEM image of the 2L-1L transition region (bottom red square in (**h**)). **m**, Scattered electron intensity profile (along the line in (**l**)) and its corresponding cross-section depicted through a ball model.

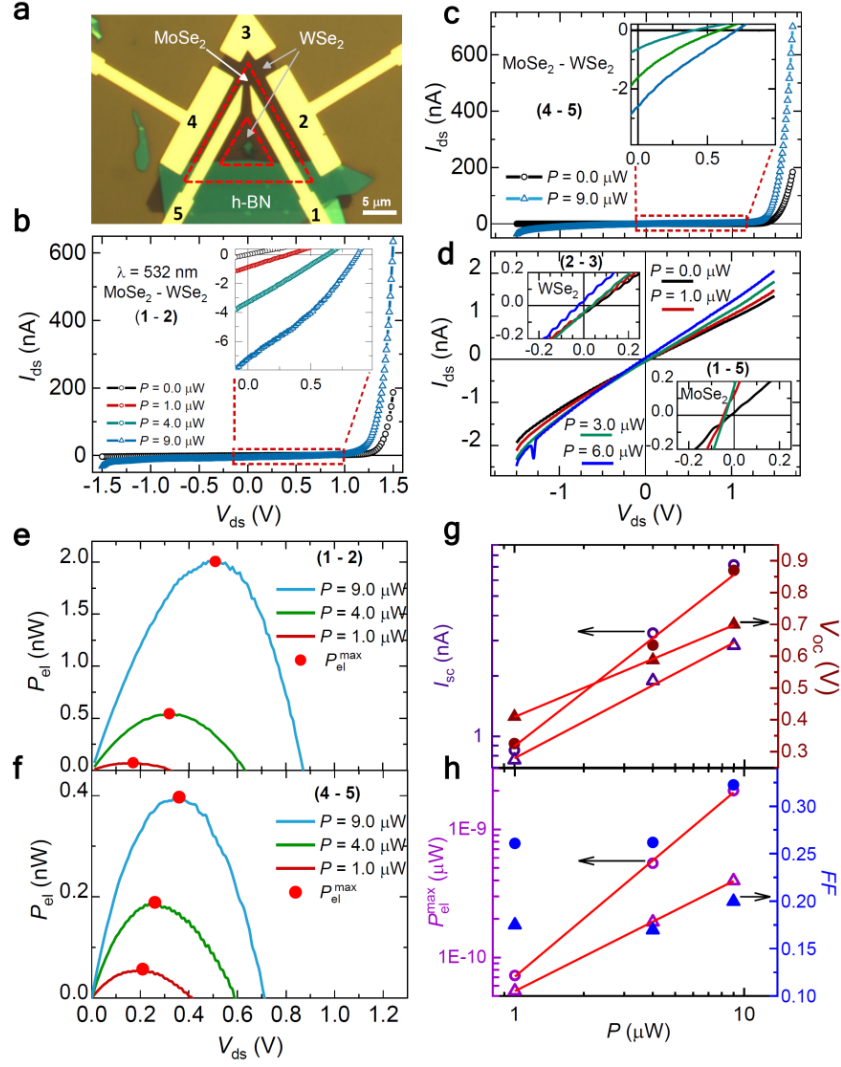

**Figure 3. Photovoltaic response across bilayer WSe$_2$-MoSe$_2$ junctions. a**, Optical image of a bilayer MoSe$_2$/WSe$_2$ bi-junction, red triangles indicate the junctions positions, gold electrical contacts are in yellow and h-BN [35] used as an insulating bridge to reach the inner domains appears in green. The response of the MoSe$_2$ domain was characterized *via* two-terminal measurements using the contacts 1-5, while the outer WSe$_2$ domain through contacts 3-4 and 2-3. The transport across the junctions, particularly under laser illumination ($\lambda$ = 532 nm), was probed through contacts 4-5 and 1-2. Notice that these pairs of contacts have distinct separations. **b**, **c**, are drain to source current $I_{ds}$ as a function of the bias voltage $V_{ds}$, for $I_{ds}$ flowing through contacts 1-2 and 5-4, respectively. Both show diode-like responses (PN-junction) that are enhanced by illumination. The insets are amplified scale plots showing short-circuit currents $I_{sc}$ (for $V_{ds}$=0) resulting from the photovoltaic effect. **d**, $I_{ds}$ vs. $V_{ds}$ within individual domains measured through contacts 3-4 (WSe$_2$-main panel and top inset) and through contacts 1-5 (MoSe$_2$-bottom inset). Notice the absence of a diode like response and the very small $I_{sc}$ values, i.e. in the order of just 0.1 nA. **e**, **f**, photo-generated electrical power $P_{el}$ = $I_{ds}$ x $V_{ds}$, under several illumination powers $P$, measured between leads 1-2 and 4-5, respectively. Red dots indicate the corresponding maxima of $P_{el}$ ($P_{el}^{max}$), which scale with the illuminated area despite the sharpness of the junctions. **g**, Short-circuit current $I_{sc}$ (dark blue symbols) and open-circuit voltage $V_{oc}$ (brown symbols), where circles (triangles) are the values extracted for the junction between leads 1-2 (4-5). The $I_{sc}$ linear fit (red line) displays a power law dependence $I_{sc} \propto P^{0.6}$. $V_{oc}$ follows the characteristic semi-logarithmic dependence on $P$ (red line linear fit). **h**, $P_{el}^{max}$ (violet hollow symbols) and fill factors $FF = P_{el}^{max}/ (I_{sc} \times V_{oc})$ (blue filled symbols) as functions of $P$. Circles and triangles depict data from the junctions between leads 1-2 and 4-5, respectively. Red lines are linear fits $P_{el}^{max} \propto P^{\alpha}$, with $\alpha = 1.5$ (junction 1-2) and $\alpha = 0.9$ (junction 4-5).

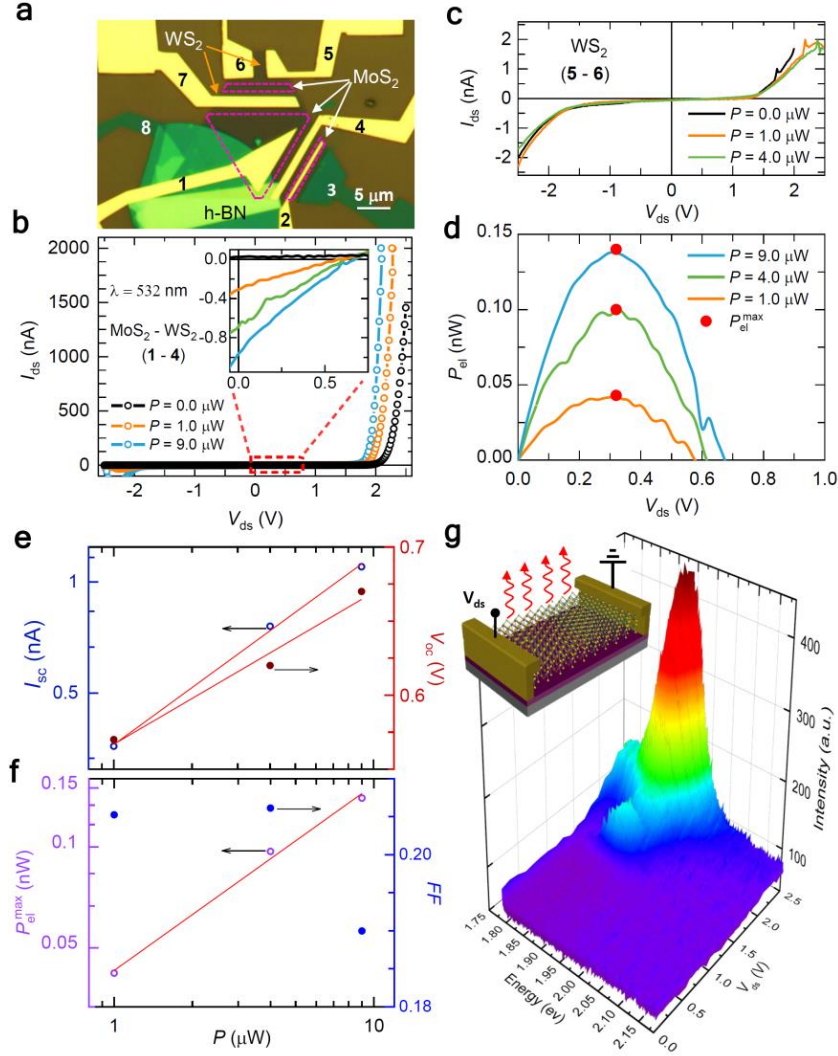

**Figure 4. Optoelectronic response of bilayer MoS$_2$-WS$_2$ junctions. a**, Optical image of a bilayer MoS$_2$/WS$_2$ three-junctions heterostructure, the junctions are highlighted by the magenta dashed lines, gold pads electrical contacts are in yellow. The optoelectronic response of the outer WS$_2$ domain was characterized through contacts 5-6; while the transport across the junction was probed through contacts 1-4. **b**, $I_{ds}$ vs. $V_{ds}$ measured across the junction (1-4), for different laser illumination powers ($\lambda$ = 532 nm). Inset: magnified scale within the dashed rectangle (around $V_{ds}$ =0) showing significant short-circuit current $I_{sc}$ resulting from the photovoltaic effect. **c**, $I_{ds}$ vs. $V_{ds}$ for currents flowing between contacts 5-6 (WS$_2$ domain), showing a non-linear *I-V* characteristic due to the Schottky barriers at the electrical contacts. Upon illumination the diode current across the junction is 10$^3$ times higher than the non-linear current flowing across a single domain. **d**, Photo-generated electrical power $P_{el}$ under different illumination powers $P$ for currents flowing between leads 1-4. Red dots indicate the corresponding $P_{el}^{max}$. **e**, Short-circuit current $I_{sc}$ (blue hollow symbols) and open-circuit voltage $V_{oc}$ (brown filled symbols) extracted for the junction (1-4). The red linear fit for $I_{sc}$ is a power law $I_{sc} \propto P^{0.5}$. The red linear fit for $V_{oc}$ follows the characteristic semi-logarithmic dependence on $P$. **f**, $P_{el}^{max}$ (violet hollow symbols) and fill factors *FF* (blue filled symbols) as functions of $P$. The red line is a linear fit yielding $P_{el}^{max} \propto P^{0.6}$. **g,** Room-temperature electro-luminescence (EL) response from a bilayer MoS$_2$-WS$_2$ lateral heterostructure.

# Supplementary Materials for

# Enhanced Optoelectronic Response in Bilayer Lateral Heterostructures of Transition Metal Dichalcogenides.

Prasana K. Sahoo[1]†*, Shahriar Memaran[2,3]†, Yan Xin[2], Tania Díaz Márquez[1], Florence Ann Nugera[1], Zhengguang Lu[2,3], Wenkai Zheng[2,3], Nikolai D. Zhigadlo[4], Dmitry Smirnov[2,3], Luis Balicas[2,3],* and Humberto R. Gutiérrez[1],*

[1] *Department of Physics, University of South Florida, Tampa, Florida 33620, USA*

[2] *National High Magnetic Field Laboratory, Florida State University, Tallahassee, FL 32310, USA*

[3] *Department of Physics, Florida State University, Tallahassee, Florida 32306, USA.*

[5] *Department of Chemistry and Biochemistry, University of Bern, 3012 Bern, Switzerland.*

† *These authors contributed equally to this work.*

\* *Corresponding authors: E-mail:* humberto3@usf.edu; balicas@magnet.fsu.edu; prasanasahoo@gmail.com

**This PDF file includes:**
  Figs. S1 to S5

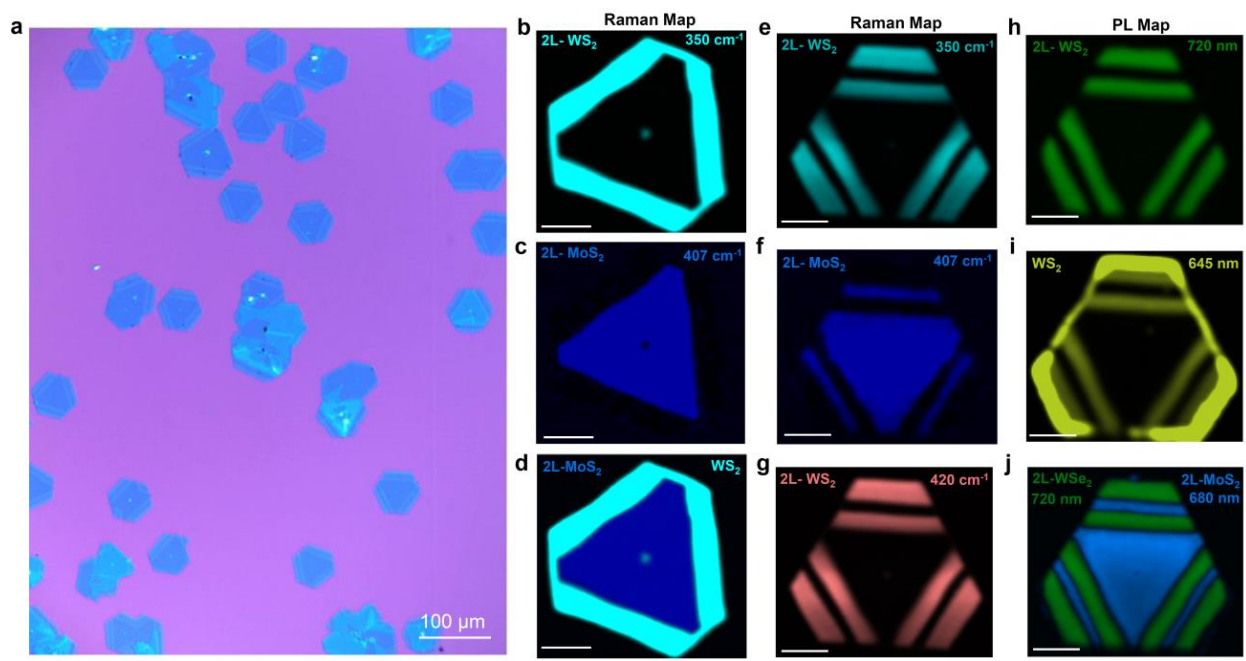

**Figure S1. Raman and photoluminescense (PL) mapping of bilayer $MoS_2$-$WS_2$ lateral heterostructures**. **a**, Low magnification optical image of a sample mainly composed of bilayer lateral heterostructures with three lateral junctions. **b-d,** Raman maps for a heterostructure with one lateral junction. **e-g**, and **h-j**, are Raman and PL maps, respectively, for a heterostructure with three lateral junction

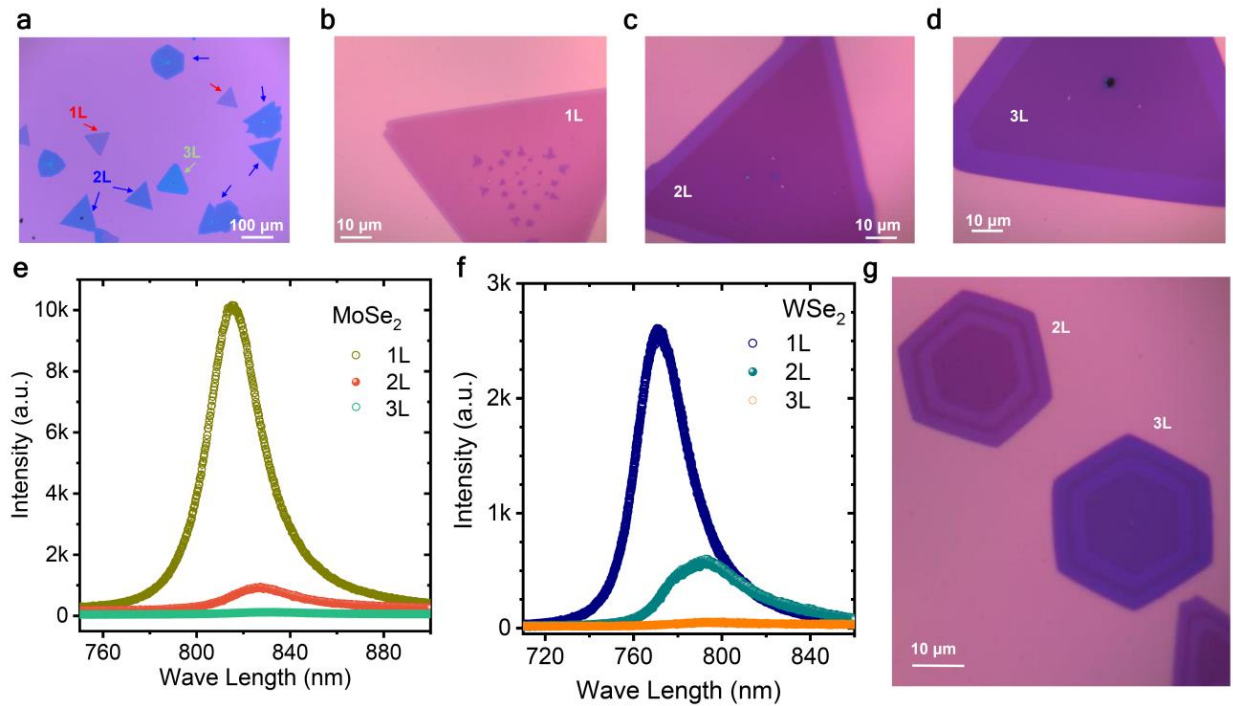

**Figure S2. Comparison between monolayer (1L), bilayer (2L) and trilayer (3L)- MoSe$_2$-WSe$_2$ lateral heterostructures**. **a**, Low magnification optical image showing the coexistence of 1L-, 2L- and 3L- lateral heterostructures within the sample indicated by the red, blue and green arrows, respectively. For the growth conditions described in the methods section, of all the heterostructures on the substrate about 90 % are 2L-heterostructures. These heterostructures have one junction and their distinct optical contrast can be used for their easy identification as shown in (**b**), (**c**) and (**d**). Photoluminescence signal from 1L, 2L and 3L individual domains of (**e**) MoSe$_2$ and (**f**) WSe$_2$ which display very distinct peak positions, peak shapes, and intensity. (**g**) Optical image from a sample with three sequential hetero-junctions, showing that the lateral pattern of distinct chemical domains is accurately reproduced either in 2L or in 3L heterostructures with approximately the same lateral domain size in each case.

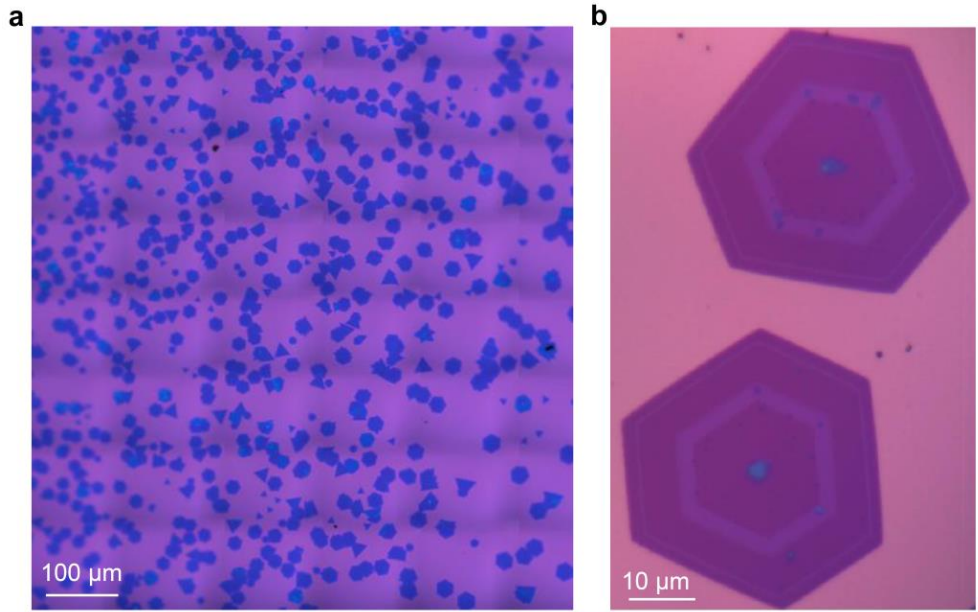

**Fig. S3. Bilayer MoSe$_2$-WSe$_2$ lateral heterostructures**. **a**, Low magnification optical image showing the homogeneous island size distribution and coverage. **b**, Magnified optical image of the islands in (**a**), in this case, each island is composed of a bilayer heterostructure with four consecutive lateral hetero-junctions, the gas switching time were intentionally varied in order to show independent control of the lateral size of each domain.

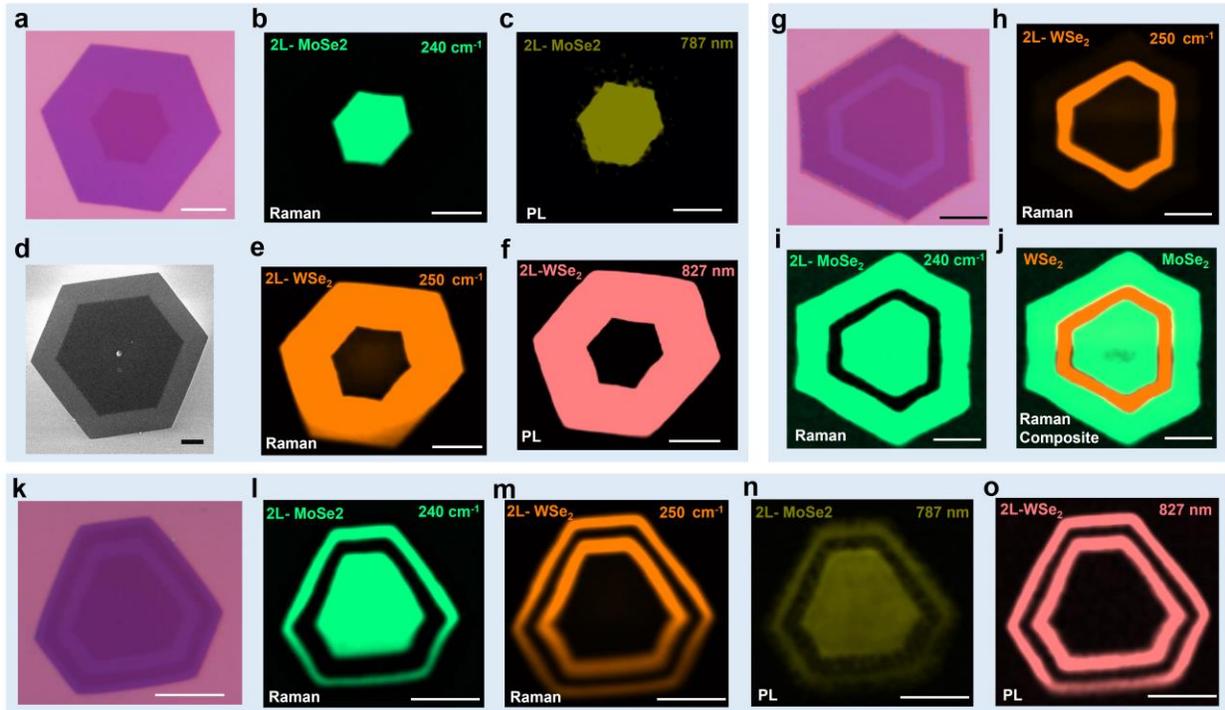

**Fig. S4. Raman and PL maps of bilayer MoSe$_2$-WSe$_2$ lateral heterostructures**. **a-f**, One-junction 2L- MoSe$_2$-WSe$_2$ heterostructure; (**a**) Optical image; (**b**) and (**e**) Raman maps; (**c**) and (**f**) PL maps and (**d**) SEM image of another one-junction heterostructure with different domain sizes. **g-j,** A 2L- MoSe$_2$-WSe$_2$ heterostructure with two junctions; (**g**) optical image; (**h-j**) Raman maps. **k-o,** A 2L- MoSe$_2$-WSe$_2$ heterostructure with three junctions; (**l**) and (**m**) are Raman map; (**n**) and (**o**) PL maps.

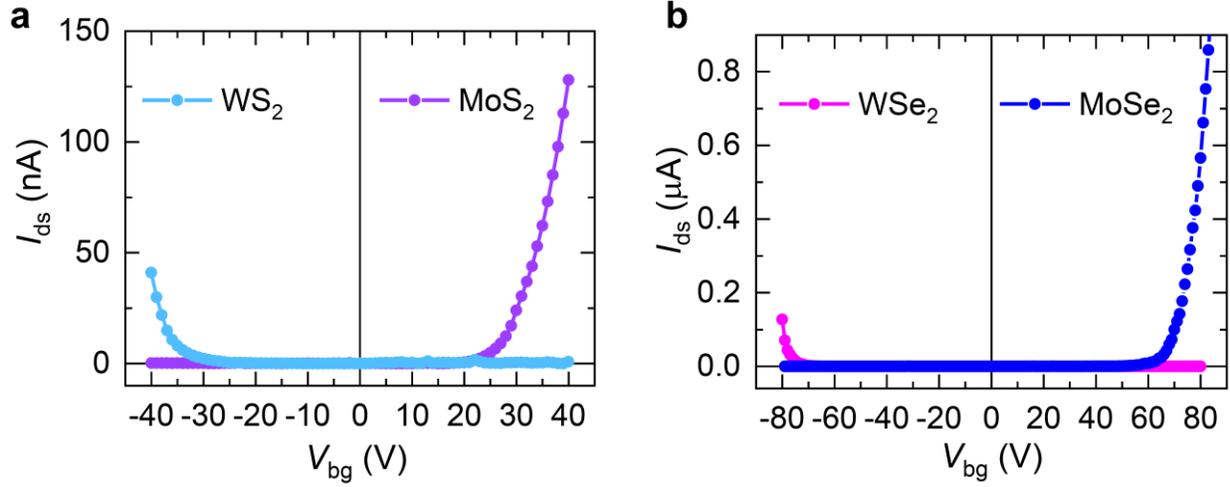

**Fig. S5. Electrical characterization of individual domains.** Drain-source current ($I_{ds}$) as a function of the back-gate voltage $V_{bg}$ for (**a**) a MoSe$_2$-WSe$_2$ and (**b**) a MoS$_2$-WS$_2$ lateral heterostructure. Both heterostructures were transferred onto a fresh SiO$_2$/Si substrate to evaluate the response of its individual domains and to minimize the current leakage through the back-gate. Notice how WS$_2$ (WSe$_2$) and MoS$_2$ (MoSe$_2$) display hole and electron doped like behavior, respectively.